\documentclass[11pt]{article}
\usepackage{graphicx}
\usepackage{algorithm,algorithmic}
\usepackage{amsmath, amssymb, amsthm}
\usepackage{url}
\usepackage{fullpage, prettyref}
\usepackage{pstricks,pst-node,pst-text}
\usepackage{boxedminipage}
\usepackage{hyperref}
\usepackage{wrapfig}
\usepackage{ifthen}

\newtheorem{theorem}{Theorem}
\newtheorem{lemma}{Lemma}
\newtheorem{claim}{Claim}
\newtheorem{corollary}{Corollary}

\newcommand{\ignore}[1]{}

\def\eps{\varepsilon}

\def\script#1{\mathcal{#1}}
\newcommand{\D}{\script{D}}

\newcommand{\junk}[1]{}
\def\E{\script{E}}

\newenvironment{proofof}[1]{\smallskip\noindent{\bf Proof of #1:}}%
        {\hfill\hspace*{\fill}$\Box$\par}
\renewenvironment{proof}[1][Proof]{\noindent \textbf{#1.}~}{\hfill$\Box$}

\def\opt{{\tt opt}}

\def\polylog{\mbox{polylog}}
\def\supp{{\tt supp}}

\begin{document}

\title{Optimal Lower Bounds for Universal and Differentially Private \\ Steiner Trees and TSPs}
\author{
Anand Bhalgat\thanks{Department of Computer and Information Science, University of Pennsylvania,
Philadelphia PA. Email: {\tt bhalgat@cis.upenn.edu}. Supported by
NSF Award CCF-0635084 and IIS-0904314.}
\and
Deeparnab Chakrabarty\thanks{Department of Computer and Information Science, University of Pennsylvania,
Philadelphia PA. Email: {\tt deepc@seas.upenn.edu}.}
\and
Sanjeev Khanna \thanks{Dept. of Computer \& Information Science, University of Pennsylvania,
Philadelphia, PA 19104. Email: {\tt sanjeev@cis.upenn.edu}. Supported in
part by NSF Awards CCF-0635084 and IIS-0904314.}
}
\date{}
\maketitle
\thispagestyle{empty}
\begin{abstract}
Given a metric space on $n$ points, an $\alpha$-approximate {\em universal} algorithm for the Steiner tree problem outputs a distribution over rooted spanning trees such that for any subset $X$ of vertices containing the root, the expected cost of the induced subtree is within an $\alpha$ factor of the optimal Steiner tree cost for $X$. An $\alpha$-approximate {\em differentially private} algorithm for the Steiner tree problem takes as input a subset $X$ of vertices, and outputs a tree distribution that induces a solution within an $\alpha$ factor of the optimal as before, and satisfies the additional property that for any set $X'$ that differs in a single vertex from $X$, the tree distributions for $X$ and $X'$ are ``close'' to each other. Universal and differentially private algorithms for TSP are defined similarly. An $\alpha$-approximate universal algorithm for the Steiner tree problem or TSP is also an $\alpha$-approximate differentially private algorithm. It is known that both problems admit $O(\log n)$-approximate universal algorithms, and hence $O(\log n)$-approximate differentially private algorithms as well. \\

We prove an $\Omega(\log n)$ lower bound on the approximation ratio achievable for the universal Steiner tree problem and the universal TSP, matching the known upper bounds. Our lower bound for the Steiner tree problem holds even when the algorithm is allowed to
output a more general solution of a distribution on paths to the root.
This improves upon an earlier $\Omega(\log n/\log \log n)$ lower bound for the universal Steiner tree problem, and an $\Omega(\log^{1/6} n)$ lower bound  for the universal TSP. The latter answers an open question in Hajiaghayi et al. \cite{HKL06}.
When expressed as a function of the size of the input subset of vertices, say $k$, our lower bounds are in fact $\Omega(k)$ for both problems, improving upon the
previously known $\log^{\Omega(1)} k$ lower bounds. We then show that whenever the universal problem has a lower bound that satisfies an additional property, it implies a similar lower bound for the differentially private version. Using this converse relation between universal and private algorithms, we establish an $\Omega(\log n)$ lower bound for the differentially private
Steiner tree and the differentially private TSP. This answers a question of Talwar \cite{Tal10}.
Our results highlight a natural connection between universal and private approximation algorithms that is likely to have other applications.
\end{abstract}

\newpage
\setcounter{page}{1}
\section{Introduction}
Traditionally, in algorithm design one assumes that
the algorithm has complete access to the input data
which it can use unrestrictedly to output the optimal, or near
optimal, solution. In many applications, however, this assumption
does not hold and the traditional approach towards algorithms needs to be revised.
For instance, let us take the problem of designing the cheapest
multicast network connecting a hub node to a set of client nodes; this
is a standard network design problem which has been studied
extensively. Consider the following two situations.
In the first setting, the actual set of clients is unknown to the algorithm,
and yet the output multicast network must be ``good for all'' possible client sets.
In the second setting, the algorithm knows the client set, however,
the algorithm needs to ensure that the output preserves the privacy
of the clients. Clearly, in both these settings, the traditional algorithms
for network design don't suffice.

The situations described above are instances of two general
classes of problems recently studied in the literature.
The first situation needs the design
of {\em universal} or {\em a-priori} algorithms;
algorithms which output solutions when parts of the input are uncertain
or unknown. The second situation needs the design of {\em differentially
private} algorithms; algorithms where parts of the input are
controlled by clients whose privacy concerns constrain the behaviour of the
algorithm. A natural question arises: how do the constraints imposed
by these classes of algorithms affect their performance?

In this paper, we study universal and differentially private algorithms for
two fundamental combinatorial optimization problems: the Steiner tree problem
and the travelling salesman problem (TSP). The network design problem mentioned
above corresponds to the Steiner tree problem.
We resolve the performance question of universal and private algorithms
for these two problems completely by giving lower bounds which match the known upper bounds.
In particular, our work resolves the open questions of Hajiaghayi et al.~\cite{HKL06} and Talwar \cite{Tal10}.
Our techniques and constructions are quite basic, and we hope these could be applicable
to other universal and private algorithms for sequencing and network design problems.

\vspace{-2mm}
\paragraph{Problem formulations.}
In both the Steiner tree problem and the TSP, we
are given a metric space $(V,c)$ on $n$ vertices with a specified root vertex $r\in V$.
Given a subset of terminals, $X\subseteq V$, we denote the cost of the
optimal Steiner tree connecting $X\cup r$ by $\opt_{ST}(X)$.
Similarly, we denote the cost of the optimal tour connecting $X\cup r$
by $\opt_{TSP}(X)$. If $X$ is known, then both $\opt_{ST}(X)$
and $\opt_{TSP}(X)$ can be approximated up to constant factors.

A {\em universal} algorithm for the Steiner tree problem, respectively the TSP, does not know the set of terminals $X$,
but must output a distribution $\D$ on rooted trees $T$, respectively tours $\sigma$, spanning all vertices of $V$.
Given a terminal set $X$, let $T[X]$ be the minimum-cost rooted subtree of $T$ which contains
$X$. Then the cost of the universal Steiner tree algorithm on terminal set $X$
is ${\bf E}_{T\leftarrow \D} [c(T[X])]$.
We say the universal Steiner tree algorithm is {\em $\alpha$-approximate},
if for all metric spaces and all terminal
sets $X$, this cost is at most $\alpha\cdot \opt_{ST}(X)$.
Similarly, given a terminal set $X$, let $\sigma_X$ denote the order in which vertices of $X$ are visited in $\sigma$,
and let $c(\sigma_X)$ denote the cost of this tour. That is,
$c(\sigma_X) := c(r,\sigma_X(1)) + \sum_{i=1}^{|X|-1} c(\sigma_X(i),\sigma_X(i+1)) + c(\sigma_X(|X|),r)$.
The cost of the universal TSP algorithm on set $X$ is ${\bf E}_{T\leftarrow \D} [c(\sigma_X)]$,
and the approximation factor is defined as it is for the universal Steiner tree algorithm.

\def\T{\mathcal T}
A {\em differentially private} algorithm for Steiner trees and TSPs, on the other hand, knows the set
of terminals $X$; however, there is a restriction on the solution that it can output. Specifically,
a differentially private algorithm for the Steiner tree problem with privacy parameter $\eps$, returns
on any input terminal set $X$ a distribution $\script{D}_X$ on trees spanning $V$,
with the following property. Fix any set of trees $\T$, and let
 $X'$ be any terminal set such that the symmetric difference of $X'$ and $X$ is exactly one vertex. Then,
$$ \Pr_{T\leftarrow \script{D}_{X'}}[T \in \T] \cdot\exp(-\eps) ~\le~ \Pr_{T\leftarrow \script{D}_{X}}[T \in \T]  ~\le~\Pr_{T\leftarrow \script{D}_{X'}}[T \in \T] \cdot \exp(\eps) $$
\noindent
The cost of the algorithm on set $X$ is ${\bf E}_{T\leftarrow \script{D}_X} [c(T[X])]$ as before, and the approximation factor is defined
as that for universal trees. Differentially private algorithms for the TSP are defined likewise.
To gain some intuition as to why this definition preserves privacy,
suppose each vertex is a user and controls a bit which reveals its identity as a terminal or not.
The above definition ensures that even if a user changes its identity, the algorithm's behaviour does not change by
much, and hence the algorithm does not leak any information about the user's identity. This notion of privacy is
arguably the standard and strongest notion of privacy in the literature today; we point the reader to \cite{Dwo06}
for an excellent survey on the same. We make two simple observations; (a) any universal algorithm is a differentially private
algorithm with $\eps = 0$, (b) if the size of the symmetric difference in the above definition is $k$ instead of $1$,
then one can apply the definition iteratively to get $k\eps$ in the exponent.

For the Steiner tree problem,
one can consider another natural and more general solution space for universal and private algorithms, where
instead of returning a distribution on trees spanning $V$, the algorithm returns a distribution $\D$ on collections
of paths $P := \{p_v:v\in V\}$, where each $p_v$ is a path from $v$ to the root $r$.
Given a single collection $P$, and a terminal set $X$, the cost of the solution is $c(P[X]) := c(\left(\bigcup_{v\in X} E(p_v)\right)$,
where $E(p_v)$ is the set of edges in the path $p_v$.
The cost of the algorithm on set $X$ is ${\bf E}_{P\leftarrow \D}[c(P[X])]$.
Since any spanning tree induces an
equivalent collection of paths, this solution space is more expressive, and as such,
algorithms in this class may achieve stronger performance guarantees.
Somewhat surprisingly, we show that this more general class of
algorithms is no more powerful than algorithms that are restricted to output a
spanning tree.


\vspace{-1mm}
\subsection{Previous Work and Our Results.}\label{sec:prevwork}
A systematic study of universal algorithms was initiated by Jia et al. \cite{JLNRS05},
who gave an $O(\log^4 n/\log \log n)$-approximate universal algorithms for both the
Steiner tree problem and the TSP. Their algorithm is in fact deterministic and returns a single
tree. Gupta et al. \cite{GHR06} improved the TSP result by giving a single tour which is
$O(\log^2 n)$-approximate.
As noted by \cite{JLNRS05}, results of \cite{Bartal,FRT03} on probabilistically embedding general metrics
into tree metrics imply randomized $O(\log n)$-approximate universal algorithms for these
problems (see Appendix \ref{sec:prelims} for details).

Jia et al.~\cite{JLNRS05} observe that a lower bound for online Steiner tree 
algorithms implies a lower bound for universal Steiner tree algorithms; thus, 
following the result of Imase and Waxman \cite{IW}, one obtains a
lower bound of $\Omega(\log n)$ for any universal Steiner tree algorithm.
It is not hard to see that the \cite{IW} lower bound also holds for 
algorithms returning a collection of vertex-to-root paths.
Jia et al.~\cite{JLNRS05} 
explicitly leave lower bounds for the universal
TSP as an open problem.
Hajiaghayi et al. \cite{HKL06} make progress on this by showing an
$\Omega\left(\sqrt[6]{\log n/\log\log n}\right)$ lower bound for universal TSP; this
holds even in the two dimensional Euclidean metric space.
\cite{HKL06} conjectured that for general metrics the lower bound should be $\Omega(\log n)$;
in fact, they conjectured this for the shortest path metric of a constant degree expander.

When the metric space has certain special properties (for instance if it is the Euclidean metric in constant dimensional space),
Jia et al. \cite{JLNRS05} give an improved universal algorithms for both Steiner tree and TSP,
which achieves an approximation factor of $O(\log n)$ for both problems.
Furthermore, if the size of the terminal set $X$ is $k$, their approximation factor
improves to $O(\log k)$ -- a significant improvement when $k\ll n$. This leads to the question whether universal algorithms
exist for these problems whose approximation factors are a non-trivial function of $k$ alone.
A $k$-approximate universal Steiner tree algorithm is trivial;
the shortest path tree achieves this factor. This in turn implies a $2k$-approximate
universal TSP algorithm. Do either of these problems admit an $o(k)$-approximate algorithm?
The constructions of \cite{IW} achieving a lower bound of $\Omega(\log n)$ for universal Steiner tree 
require terminal sets that are of size
$n^{\Omega(1)}$, and do not rule out the possibility of an $O(\log k)$-approximation in general.
In fact, for many network optimization problems, an initial
$\polylog(n)$ approximation bound was subsequently improved to a $\polylog(k)$ approximation (e.g.,
sparsest cut \cite{LR88,LLR95}, asymmetric $k$-center \cite{PV98,Archer01}, and
more recently, the works of Moitra et al. \cite{Moitra09,ML10} on vertex sparsifiers imply
such a result for other many cut and flow problems). It is thus conceivable that a $\polylog(k)$-approximation
could be possible for the universal algorithms as well. 
\smallskip

\indent
{\em 
We prove $\Omega(\log n)$ lower bounds for 
the universal TSP and the Steiner tree problem, 
even when the algorithm returns vertex-to-root 
paths for the latter (Theorems \ref{thm:univtsp} and \ref{thm:univst-paths}). Furthermore, the size of the terminal 
sets in our lower bounds is $\Theta(\log n)$, ruling out 
any $o(k)$-universal algorithm for either of these 
problems.} (Very recently, we were made aware of independent work by Gorodezky et al. \cite{GKSS10}
who obtained similar lower bounds for the universal TSP problem.
We make a comparison of the results of our work and theirs at the end of this subsection.)

%


\medskip
\noindent
{\em Private vs universal algorithms.}
The study of differentially private algorithms for combinatorial optimization problems is much newer, and the paper by
Gupta et al. \cite{GLMRT10} gives a host of private algorithms for many optimization problems.
Since any universal algorithm is a differentially private algorithm with $\eps=0$, the above stated upper bounds
for universal algorithms hold for differentially private algorithms as well.
For the Steiner tree problem and TSP, though, no better differentially  private algorithms are known.
Talwar, one of the authors of \cite{GLMRT10}, recently posed an open question whether
a private $O(1)$-approximation exists for the Steiner tree problem, even if the algorithm is allowed to use a more general solution space, namely,
return a collection of vertex-to-root paths, rather than Steiner trees \cite{Tal10}.

We observe that a simple but useful converse relation holds between universal and private algorithms:
``strong" lower bounds for universal algorithms implies lower bounds for differentially private algorithms.
More precisely, suppose we can show that for any universal algorithm for the Steiner tree problem/TSP, there exists
a terminal set $X$, such that the probability that a tree/tour drawn from the distribution has cost
less than $\alpha$ times the optimal cost is $\exp(- \eps |X|)$ for a certain constant $\eps$. {\em Then we get
an $\Omega(\alpha)$ lower bound on the performance of any $\eps$-differentially private algorithm for these problems.} (Corollary \ref{cor:st}).
Note that this is a much stronger statement than merely proving a lower bound on the expected cost
of a universal algorithm. The expected cost of a universal algorithm may be $\Omega(\alpha)$, for instance,
even if it achieves optimal cost with probability $1/2$, and $\alpha$ times the optimal cost with probability $1/2$. 
In fact, none previous works mentioned above~\cite{IW,JLNRS05,HKL06} imply strong lower bounds.
The connection between strong lower bound on universal algorithms and lower bounds for
differentially private algorithms holds for a general class of problems, and may serve as a useful tool
for establishing lower bounds for differentially private algorithms (Section \ref{sec:univpriv}).


In contrast to previous work,
all the lower bounds we prove for universal Steiner trees and TSP are strong in the sense defined above.
Thus, as corollaries, {\em we get lower bounds of $\Omega(\log n)$ on the performance of differentially private algorithms
for Steiner tree and TSP. Since the lower bound for Steiner trees holds even when the algorithm returns a collection of paths, this answers the question of Talwar \cite{Tal10} negatively.} (Corollaries \ref{cor:st} and \ref{cor:tsp}).

%

The metric spaces for our lower bounds on universal Steiner tree and TSP
are shortest path metrics on constant degree Ramanujan expanders. To prove the strong lower
bounds on distributions of trees/tours, it suffices, by Yao's lemma, to construct
a distributions on terminal sets such that any fixed tree/tour pays, with high
probability, an $\Omega(\log n)$ times the optimum tree/tour's cost on
a terminal set picked from the distribution. We show that a random walk,
or a union of two random walks, suffices for the Steiner tree and the TSP case,
respectively.

\medskip
\noindent
{\em Comparison of our results with \cite{GKSS10}:}
Gorodezky et al.~\cite{GKSS10} independently obtained an $\Omega(\log n)$ lower bound for universal TSP.
Like us, the authors construct the lower bound using random walks on constant degree expanders.
Although the result is stated for deterministic algorithms, Theorem 2 in their paper implies
that the probability any randomized algorithm pays $o(\log n)$ times the optimum for a certain
subset is at most a constant. Furthermore, their result also implies an $\Omega(k)$ lower bound on the performance
of a universal TSP algorithm where $k$ is the number of terminals. 

Although \cite{GKSS10} do not address universal Steiner tree problem directly, the $\Omega(k)$ lower bound
for universal TSP implies an $\Omega(k)$
lower bound for universal Steiner tree as well, only when the algorithm returns spanning trees.
However, this doesn't work for algorithms which return collections of vertex-to-root paths.
Our result provides the first $\Omega(k)$ lower bound for the universal Steiner tree problem when the algorithm is allowed to return a collection of vertex-to-root paths. 

Furthermore, even though our proof idea is similar, our 
results are stronger since we show a ``strong'' lower bounds for the
universal problems: we prove that the probability any randomized algorithm pays $o(\log n)$ times the optimum for a certain subset is exponentially small in the size of the client set. (We state the precise technical difference in Section \ref{sec:tsp} while describing our lower bound.)
As stated above, strong lower bounds are necessary in our technique for proving privacy lower bounds.
In particular, no lower bounds  for differentially private  Steiner tree (even for weaker algorithms returning spanning trees instead of vertex-to-root paths) and TSP can be deduced from their results.

\vspace{-3mm}
\subsection{Related Work}
Although universal algorithms in their generality were first studied  by Jia et al.\cite{JLNRS05},
the universal TSP on the plane was investigated by Platzman and Bartholdi \cite{PB89},
who showed that a certain space filling curve is an $O(\log n)$-approximate algorithm for points
on the two dimensional plane. Bertsimas and Grigni \cite{BG89} conjecture that this factor is
tight, and \cite{HKL06} makes progress in this direction,
although till this work, it was not known even for points in a general metric space. It is an interesting
open question to see if our ideas could be modified for the special metric as well.

The notion of differential privacy was developed in the regime of statistical data analysis
to reveal statistics of a database without leaking any extra information of individual entries;
the current adopted definition is due to Dwork et al.\cite{DMNS06}, and since its definition
a large body of work has arisen trying to understand the strengths and limitations of this concept.
We point the reader to excellent surveys by Dwork and others \cite{Dwo06,Dwo08,DS09} for
a detailed treatment. Although the notion of privacy arose in the realm of databases, the concept
is more universally applicable to algorithms where parts of the inputs are controlled by
privacy-concerned users. Aside from the work of Gupta et al.\cite{GLMRT10} on various combinatorial optimization
problems, algorithms with privacy constraints have been developed for other problems such as
computational learning problems \cite{KLNRS08}, geometric clustering problems \cite{FFKN09},
recommendation systems \cite{MM09}, to name a few.


\smallskip
\noindent
{\bf Organization.}
In Section~\ref{sec:Steiner_tsp}, we establish an $\Omega(\log n)$ lower bound for the
universal Steiner tree problem and the universal TSP. As mentioned above, the lower bound for the
Steiner tree problem is for a more general class of algorithms which return a collection of paths
instead of a single tree.
The lower bound established are strong in the sense defined earlier, and thus give
an $\Omega(\log n)$ lower bound for private Steiner tree as well as private
TSP. We formalize the connection between strong lower bounds for universal problems
and approximability of differentially private variants in Section~\ref{sec:univpriv}.
Finally, for sake of completeness, we provide in Appendix~\ref{sec:upper_bounds}
a brief description of some upper bound results that follow implicitly from earlier works.

\vspace{-2mm}
\section{Lower Bound Constructions}
\label{sec:Steiner_tsp}
The metric spaces on which we obtain our lower bounds are shortest path metrics of
expander graphs. Before exhibiting our constructions, we state a few known results
regarding expanders that we use. 
An $(n,d,\beta)$ expander is a $d$ regular, $n$ vertex graph with the second largest eigenvalue of
its adjacency matrix $\beta < 1$. The girth $g$ is the size of the smallest cycle and the diameter
$\Delta$ is the maximum distance between two vertices.
A $t$-step random walk on an expander picks a vertex uniformly at
random, and at each step moves to a neighboring vertex uniformly at random.

\begin{lemma}\label{lem:expanderexistence}\cite{LPS88}
For any constant $k$, there exist $(n,d,\beta)$ expanders, called
Ramanujan graphs, with $d \ge k$, $\beta \le \frac{2}{\sqrt{d}}$, girth $g=\Theta(\log n/\log d)$,
and diameter $\Delta = \Theta(\log n/\log d)$.
\end{lemma}

\vspace{-3mm}
\begin{lemma}\label{lem:alphabet}(Theorem 3.6, \cite{HLW06})
Given an $(n,d,\beta)$ expander, and a subset of vertices $B$ with $|B|=\alpha n$,
the probability that a $t$-step random walk remains completely inside $B$ is at most
$(\alpha + \beta)^t$.
\end{lemma}

\vspace{-3mm}
\begin{lemma}\label{lem:alphabet2}(Follows from Theorem 3.10, \cite{HLW06})
Given an $(n,d,\beta)$ expander, a subset of vertices $B$ with $|B|=\alpha n$, and any $\gamma, 0\le \gamma \le 1$,
the probability that a $t$-step random walk visits more than $\gamma t$ vertices in $B$
is at most  $2^t\cdot (\alpha + \beta)^{\gamma t}$.
\end{lemma}
\vspace{-4mm}
\subsection{Steiner Tree Problem}
We consider a stronger class of algorithms that are allowed to
return a distribution $\D$ on collections of paths $P := \{p_v:v\in V\}$,
where each $p_v$ is a path from $v$ to the root $r$. As stated in the introduction, this class
of algorithms captures as a special case algorithms that simply return a
distribution on collection of spanning trees, since the latter induces a collection of paths.
We prove the following theorem.

\begin{theorem}\label{thm:univst-paths}
For any constant $\eps > 0$ and for large enough $n$, there exists a metric space $(V,c)$ on $n$ vertices
such that for any distribution $\D$ on collections
of paths, there is a terminal set $X$ of size $\Theta(\log n)$, such that
\begin{equation}\label{eq:lb}
\Pr_{P\leftarrow \D}\left[c(P[X]) = o\left(\frac{\log n}{1+\epsilon}\right)\opt_{ST}(X)\right] \le \frac{1}{2}\exp(-\eps |X|)
\end{equation}
\end{theorem}

\noindent
At a high-level, the idea underlying our proof is as follows.
We choose as our underlying graph a Ramanujan graph $G$, and consider the shortest
path metric induced by this graph. We show that for any fixed collection $P$
of vertex-to-root paths, a terminal set generated by a random walk $q$
of length $\Theta(\log n)$ in $G$ has the following property with high probability:
the edges on $q$ frequently ``deviate'' from the paths in the collection $P$. These deviations
can be mapped to cycles in $G$, and the high-girth property is then used to establish that
the cost of the solution induced by $P$ is $\Omega(\log n)$ times the optimal
cost. Before proving Theorem~\ref{thm:univst-paths}, we establish the following corollaries of it.

\begin{corollary}\label{cor:st}
(a) There is no $o(\log n)$-approximate universal Steiner tree algorithm. (b) There is no $o(k)$-approximate universal Steiner tree
algorithm where $k$ is the size of the terminal set. (c) For any $\eps > 0$, there is no $o(\log n/(1+\eps))$-approximate private algorithm with privacy parameter $\eps$.
\end{corollary}
\begin{proof}
The proofs of (a) and (b) are immediate by fixing $\eps$ to be any constant.
The universal algorithm pays at least $\Omega(\log n)$ times the optimum with high probability,
thus giving a lower bound of $\Omega(\log n)$ on the expected cost.
To see (c), consider a differentially private algorithm $\script{A}$ with privacy parameter $\eps$.
Let $\script{D}$ be the distribution on the collection of paths returned by $\script{A}$ when the terminal set is $\emptyset$.
Let  $X$ be the subset of vertices corresponding to this distribution in Theorem \ref{thm:univst-paths}.
Let $\script{P} := \{P: c(P[X]) = o( \frac{\log n}{1+\epsilon})\cdot \opt_{ST}(X)\}$; we know $\Pr_{P\leftarrow \script{D}}[P\in \script{P}] \le \frac{1}{2}\exp(-\eps |X|)$.
Let $\script{D}'$ be the distribution on the collection of paths returned by $\script{A}$ when the terminal set is $X$. By the definition of $\eps$-differential privacy, we know that $\Pr_{P\leftarrow \script{D}'} [P \in \script{P}] \le \exp(\eps\cdot |X|)\cdot \left( \frac{1}{2}\exp(-\eps |X|)\right)\le 1/2$.
Thus with probability at least $1/2$, the differentially private algorithm returns a collection of path of cost
at least $\Omega\left(\frac{\log n}{1+\epsilon}\right)\cdot \opt_{ST}(X)$, implying the lower bound.
\end{proof}

Note that the statement of Theorem \ref{thm:univst-paths} is much stronger than what is needed to prove
the universal lower bounds.
The proof of part (c) of the above corollary illustrates our observation that showing strong lower bounds for universal problems
imply lower bounds for privacy problems. This holds more generally, and we explore this more in Section \ref{sec:univpriv}.
We now prove of Theorem \ref{thm:univst-paths}.

\begin{proofof}{Theorem \ref{thm:univst-paths}}
Consider an $(n,d,\beta)$ expander as in Lemma \ref{lem:expanderexistence} with degree $d \ge 2^{K(1+\epsilon)}$,
where $K$ is a large enough constant.
The metric $(V,c)$ is the shortest path metric induced by this expander.
The root vertex $r$ is an arbitrary vertex in $V$.

We now demonstrate a distribution $\D'$ on terminal sets $X$
such that $\eps |X| \le C_0\log n$, for some constant $C_0$, and
for any fixed collection of paths $P$,

\begin{equation}\label{eq:yao}
\Pr_{X\leftarrow \D'}\left[c(P[X]) = o\left(\frac{\log n}{1+\eps}\right)\opt_{ST}(X)\right] \le \frac{1}{2}\exp(-C_0 \log n).
\end{equation}

The lemma below is essentially similar to Yao's lemma \cite{Yao77} used for
establishing lower bounds on the performance of randomized algorithms against oblivious
adversaries.

\begin{lemma}\label{lem:yao}
Existence of a distribution $\D'$ satisfying \eqref{eq:yao} proves Theorem \ref{thm:univst-paths}.
\end{lemma}
\begin{proof}
For brevity, denote the expression in the RHS of \ref{eq:yao} by $\rho$. Let $\pi_X$ be the
probability of $X$ in the distribution $\D'$ and $\pi_P$ be the probability of collection $P$
in the distribution $\D$. Let $\E(P,X)$ denote the event $c(P[X]) = o(\log n/(1+\epsilon))\opt_{ST}(X)$. Then
\eqref{eq:yao} implies that for each $P$ in the support of $\D$,
we have $\sum_{\small X\in \supp(\D'): \E(P,X)} \pi_X \le \rho$.
Thus, $\sum_{P\in \supp(\D)} \pi_P \left(\sum_{X\in \supp(\D'): \E(P,X)} \pi_X\right) \le \rho$,
and interchanging summations, $\sum_{X\in \supp(\D')} \pi_X \left(\sum_{P\in \supp(\D): \E(P,X)} \pi_P\right) \le \rho$, which
implies that there exists $X\in \supp(\D')$ such that $\Pr_{P\leftarrow \D}[c(P[X]) = o(\log n/(1+\epsilon))\opt_{ST}(X)] \le \frac{1}{2}\exp(-C_0\log n) \le \frac{1}{2} \exp(-\eps |X|)$.
\end{proof}

\indent
The distribution $\D'$ is defined as follows.
Recall that the girth and the diameter of $G$ are denoted by $g$ and $\Delta$ respectively,  and both are $\Theta\left(\frac{\log n}{\log d}\right)$.
Consider a random walk $q$ of $t$-steps in $G$, where $t = g/3$, and let
$X$ be the set of distinct vertices in the random walk. This defines the distribution
on terminal sets. Note that each $X$ in the distribution has size $|X|=O(\log n/\log d)$.
We define $C_0$ later to be a constant independent of $d$, and thus since $d$ is large
enough, $\eps|X| \le C_0\log n$.

Fix a collection of paths $P$. Since we use the shortest path metric of $G$, we may assume
that $P$ is a collection of paths in $G$ as well.
Let $(v,v_1)$ be the first edge on the path $p_v$, and let $F := \{(v,v_1): v\in V\}$ be the collection of all these first edges.
The following is the crucial observation which gives us the lower bound.
Call a walk $q = (u_1,\ldots,u_t)$
on $t$ vertices {\em good} if at most $t/8$ of the edges of the form $(u_i,u_{i+1})$ are in $F$,
and it contains at least $t/2$ distinct vertices.

\begin{lemma}\label{lem:girth}
Let $q$ be a good walk of length $t = g/3$ and let $X$ be the set of distinct vertices in $q$.
Then $c(P[X]) = \Omega(|X|g)$.
\end{lemma}
\begin{proof}
Let $X'$ be the vertices in $X$ which do not traverse edges in $F$ in the random walk $q$.
Thus $|X'| \ge |X| - 2t/8 \ge |X|/2$. We now claim that $c(P[X']) \ge |X'|g/3$ which
proves the lemma. For every $u\in X'$, let $p'_u$ be the first
$g/3$ edges in the path $p_u$ (if $p_u$'s length is smaller than $g/3$, $p'_u = p_u$).
All the $p'_u$'s are vertex disjoint: if $p'_u$ and $p'_v$ intersect then the union of the edges in
$p'_u$, $p'_v$ and the part of the walk $q$ from $v$ to $u$ contains a cycle of length at most
$g$ contradicting that the girth of $G$ is $g$. Thus, $c(P[X'])$, which is at least $c(\bigcup_{u\in X'} p'_u) \ge |X'|g/3 \ge |X|g/6$.
\end{proof}

\noindent

Call the set of edges $F$ {\em bad}; note that the number of bad edges is at most $n$.
Lemma \ref{lem:randexpand},
which we state and prove below,
implies that the probability a $t$-step random walk is good is at least
$(1 - d^{-\Omega(t)})$. Observe that this expression is $(1 - \exp(- C_0 \log n))$ for
a constant $C_0$ independent of $d$.
Furthermore, whenever $q$ is a good walk, the set of distinct vertices $X$ in
$q$ are at least $t/2$ in number; therefore $\opt_{ST}(X) \le t + \Delta = \Theta(|X|)$
since one can always connect $X$ to $r$ by travelling along $q$ and then connecting to $r$.
On the other hand, Lemma~\ref{lem:girth} implies that $c(P[X]) = \Omega(|X|g) = \Omega(\frac{\log n}{\log d})\cdot \opt_{ST}(X) = \Omega(\frac{\log n}{1+\eps})\cdot\opt_{ST}(X)$, by our choice of $d$.
This gives that
$$\Pr_{X\leftarrow \D'}[c(P[X])\le o\left(\frac{\log n}{ 1+\eps} \right)\opt_{ST}(X)] \le \frac{1}{2} \exp(-C_0\log n)$$
where $C_0$ is independent of $d$.
Thus,
$\D'$ satisfies \eqref{eq:yao}, implying, by Lemma \ref{lem:yao}, Theorem \ref{thm:univst-paths}.
\end{proofof}
\def\E{\script{E}}
\begin{lemma}\label{lem:randexpand}
Let $G$ be an $(n,d,\beta)$ expander where $d$ is a large constant
$(\ge 2^{100}, \mbox{say})$ and
$\beta = \frac{2}{\sqrt{d}}$. Suppose we mark an arbitrarily chosen subset of $n$ edges in $G$ as bad. Then the probability that a $t$ step random walk contains at most $t/8$ bad edges
and covers at least $t/2$ distinct vertices is at least $(1 - d^{-\Omega(t)})$.
\end{lemma}
\begin{proof}
Let $\E_1$ be the event that a $t$ step random walk contains fewer than $t/2$ distinct vertices, and
let $\E_2$ be the event that a $t$ step random walk  contains at least $t/8$ bad edges.
We bound these probabilities separately.

\begin{claim}\label{lem:ev1}
$\Pr[\E_1] = d^{-\Omega(t)}$.
\end{claim}
\begin{proof}
Partition $V$ arbitrarily into $\ell = \frac{t\sqrt{d}}{2}$ sets of size $\frac{2n}{t\sqrt{d}}$ vertices each.
$\Pr[\E_1]$ can be bounded by the probability that a $t$ step random walk visits fewer than $t/2$ of these sets.
Since any fixed set of $t/2$ sets contains at most $\alpha n := n/\sqrt{d}$ vertices, by Lemma \ref{lem:alphabet}, the probability
that a $t$ step random walk remains inside the union of these sets
is at most $(3/\sqrt{d})^t$. By a union bound over all possible choices of
$t/2$ sets, we get
$$\Pr[\E_1] \le {t\sqrt{d}/2 \choose t/2} \cdot (3/\sqrt{d})^t \le (4\sqrt{d})^{t/2} (3/\sqrt{d})^t \le (3/\sqrt{d})^{t/4} \le d^{-t/12}.$$
The last two inequalities follows since $d$ is large enough.
\end{proof}

We now bound $\Pr[\E_2]$.
Call a vertex {\em bad} if more than $\sqrt{d}$ incident edges are bad.
Vertices and edges which are not bad are called good.
The set of bad vertices,
denoted by $B$, has size at most $\alpha n \le 2n/\sqrt{d}$.
%
%
%
Now consider the modification to the random walk which terminates when it visits at least $15t/16$ good vertices and at least
$t$ vertices in all. We define two bad events for the modified random walk experiment.
We say event $\E_{21}$ occurs if is the length of the modified walk is more than length $t$, and that
event $\E_{22}$ occurs if the modified walk traverses fewer than $7t/8$ good edges.

\begin{claim}\label{lem:ev2}
$\Pr[\E_2] \le \Pr[\E_{21}] + \Pr[\E_{22}]$.
\end{claim}
\begin{proof}
Observe that any walk of length exactly $t$ which occurs with non-zero probability in the modified random walk,
also occurs with the same probability in the original random walk. If a walk has at least $7t/8$ good edges,
then the set of these walks form a subset of walks in the original experiment in which $\E_2$ does not occur. So,
$\Pr[\lnot \E_2] \ge \Pr[\lnot \E_{21} \wedge \lnot \E_{22}] \ge 1 - (\Pr[\E_{21}] + \Pr[\E_{22}])$.
\end{proof}
\noindent
\begin{claim}\label{claim:ev2122}
(a) $\Pr[\E_{21}] \le d^{-\Omega(t)}$.
(b) $\Pr[\E_{22}] \le d^{-\Omega(t)}$.
\end{claim}
\begin{proof}
Part (a) follows from Lemma \ref{lem:alphabet2} where $B$ is the set of bad vertices having size at most $2n/\sqrt{d}$.
Thus the probability a random walk of length $t$ contains more than $t/16$ bad vertices is at most $2^t\cdot(4/\sqrt{d})^{t/16} \le d^{-t/64}$,
since $d$ is large enough.

For part (b), define random variables $X_1,\ldots,X_\ell$, where $\ell=15t/16$, as follows.
Each $X_i$ takes a value
when the random walk visits the $i$th good vertex $v$ on its path. Let $f$ be the fraction of good edges incident on $v$.
Since $v$ is good, we know $f \ge (1-1/\sqrt{d})$. Now, from $v$ if the random walk traverses a bad edge, set $X_i = 0$.
If the random walk traverses a good edge, toss a coin which is heads with probability $(1-\frac{1}{\sqrt{d}})/f \le 1$, and set $X_i = 1$
if the coin falls heads, else set $X_i = 0$.
Firstly, note that the probability $\Pr[X_i = 1] = f\cdot(1-1/\sqrt{d})/f = (1 - 1/\sqrt{d})$.
Secondly, note that the number of good edges traversed is at least $\sum_{i=1}^r X_i$.
Finally, and most crucially, note that the $X_i$'s are independent since the coin tosses are independent at each $i$.
Since $d$ is large enough, we get
$$\Pr[\E_{22}] \le \Pr[\sum_{i=1}^\ell X_i < 7t/8] \le 2^{15t/16} \left(\frac{1}{\sqrt{d}}\right)^{t/16} \le d^{-t/64}$$

\end{proof}

\vspace{-1mm}
\noindent
To complete the proof of Lemma \ref{lem:randexpand}, note that the probability a $t$ step random walk contains at most $t/8$
bad edges and consists of at least $t/2$ distinct vertices is $\Pr[\lnot \E_1 \wedge \lnot \E_2] \ge 1 - (\Pr(\E_1)  + \Pr[\E_2]) \ge 1 - d^{-\Omega(t)}$,
from Claims \ref{lem:ev1}, \ref{lem:ev2} and \ref{claim:ev2122}. \end{proof}

\vspace{-3mm}
\subsection{Traveling Salesman Problem}\label{sec:tsp}

We now show an $\Omega(\log n)$ lower bound for the traveling salesman problem.
In contrast to our result for the Steiner tree problem, the TSP result is
slightly weaker result in that it precludes the existence of $o(\log n)$-approximate private algorithms
for arbitrarily small constant privacy parameters only.

We remark here that a lower bound for universal TSP implies a similar lower bound for any universal
Steiner tree algorithm which returns a distribution on spanning trees.
However, this is not the case when the algorithm returns a collection of paths; in particular, our next theorem below does not imply Theorem \ref{thm:univst-paths} even in a weak sense, that is, even if
we restrict the parameter $\eps$ to be less than the constant $\eps_0$  (see Appendix \ref{sec:prelims} for details).

\begin{theorem}\label{thm:univtsp}
There exists a metric space $(V,c)$ and a constant $\eps_0$,
such that for any distribution $\D$ on tours $\sigma$ of $V$,
there exists a set $X\subseteq V$ of size $\Theta(\log n)$ such that
$$\Pr_{\sigma\leftarrow \D}[c(\sigma_X) = o(\log n)\cdot \opt_{TSP}(X)] \le \frac{1}{2}\exp(-\eps_0 |X|)$$
\end{theorem}

At a high level, the idea as before is to choose
as our underlying graph a Ramanujan graph $G$, and consider the shortest
path metric induced by this graph. We show that for any fixed permutation $\sigma$ of vertices,
with high probability a {\em pair} of random walks, say $q_1, q_2$, has the property that they frequently alternate with respect to $\sigma$. Moreover, with high probability, every vertex on $q_1$ is $\Omega(\log n)$ distance from every vertex in $q_2$. The alternation along with
large pairwise distance between vertices of $q_1$ and $q_2$ implies that on input set
defined by vertices of $q_1$ and $q_2$, the cost of the tour induced by
$\sigma$ is $\Omega(\log n)$ times the optimal cost.

As stated in the Introduction, Gorodezky et al. \cite{GKSS10} also consider the shortest path metric
on Ramanujan expanders to prove their lower bound on universal TSP. However, instead of taking
clients from two independent random walks, they use a single random walk to obtain their
set of  `bad' vertices. Seemingly, our use of two random walks makes the proof easier, and allows us
to make a stronger statement: the RHS in the probability claim in Theorem \ref{thm:univtsp} is
exponentially small in $|X|$, while \cite{GKSS10} implies only a constant. This is not sufficient for part (c)
of the following corollary.

As in the case of Steiner tree problem, we get the following corollaries of the above theorem.

\begin{corollary}\label{cor:tsp}
(a) There is no $o(\log n)$-approximate universal TSP algorithm. (b) There is no $o(k)$-approximate
universal TSP algorithm where $k$ is the size of the terminal set.
(c) There exists $\eps_0 > 0$ such that there is no $o(\log n)$-approximate
private algorithm with privacy parameter at most $\eps_0$.
\end{corollary}

\vspace{-2mm}
\begin{proofof}{Theorem~\ref{thm:univtsp}}
In the proof below we do not optimize for the constant $\eps_0$.
Using Lemma \ref{lem:expanderexistence}, we pick an $(n, d, \beta)$ expander of diameter $O(\log n)$,
where $d$ is a constant such that $\beta\le 1/10$. Let
$(V,c)$ be the corresponding metric space obtained via the shortest path metric and choose a vertex $r$ as the root vertex.
As in the proof of Lemma \ref{lem:yao},
it suffices to construct a distribution $\D' $ on subsets $X$ of size at most $C_0\log n/\eps_0$, for some constant $C_0$, 
such that given any permutation $\sigma$ on the vertices of $G$,
\begin{equation}\label{eq:tsp}
\Pr_{X\leftarrow  \D'} [c(\sigma_X) \le o(\log n)\opt_{TSP}(X)] \le \frac{1}{2}\exp(- C_0 \log n)
\end{equation}

We construct $\D'$ as follows. Pick a vertex uniformly
at random and perform a random walk $q_1$ for $t := \frac{\log_d n}{4}$ steps.
Let $X_1$ be the set of vertices visited in this walk.
Repeat this process independently to generate a second
walk $q_2$ and let $X_2$ be the set of vertices visited in the second random walk.
The set of vertices visited by the two walks together define our terminal set, namely,
$X = X_1 \cup X_2$. Note that $|X| \le  \frac{\log n}{2\log d} = \Theta(\log n)$.
Since the diameter of the graph is $O(\log n)$, we have $\opt_{TSP}(X) = O(\log n)$.
This defines the distribution $\D'$.

Let $\E_1$ be the event that the starting point of $q_2$ is at distance at least $3t$ from the starting point of $q_1$. Thus when the event $\E_1$ occurs, each vertex in $X_1$ is at distance at least $t$ from any vertex in $X_2$.
Note that, $\Pr[\E_1]$ is exactly the fraction of vertices in $G$ which are at distance at least $3t$ from any given vertex.
Since at most $d^{3t}( = n^{3/4})$ vertices are at a distance $3t$ from any vertex,
$\Pr[\E_1]$  is at least $(1-n^{-1/4}) = (1 - \exp(- \Omega(\log n))$.

We partition $\sigma$ into $\ell = \gamma {\log_d n}$ blocks of length $n/\ell$
each where $\gamma$ is a constant to be specified later in the proof of Claim \ref{lem:P1andP2}.
Let $\E_2$ denote the event that both $q_1$ and $q_2$ visit at least $3\ell/4$ blocks each.
The claim below shows that this event occurs with high probability.

\begin{claim}
\label{lem:P1andP2}
$\Pr_{X\leftarrow \D'}[\E_2] \ge (1-\exp(-\Omega(\log_d n))$.
\end{claim}
\begin{proof}
By symmetry, it suffices to analyze the probability of the event that $q_1$ visits fewer than $3\ell/4$ blocks.
Fix any set of $3\ell/4$ blocks, and let $B$ denote the union of these
$3\ell/4$ blocks. By Lemma~\ref{lem:alphabet}, the probability that $q_1$ remains inside $B$ is bounded by
$ \left(\beta + \frac{3}{4}\right)^{t} \leq \left( \frac{1}{10} + \frac{3}{4} \right)^{\frac{ \log_d n}{4} } = ~2^{- (C_1\log_d n)}$
for some  constant $C_1 > 0$.  
Set  $\gamma$ to be $C_1/2$.
The probability that $X_1$ visits fewer than $3\ell/4$ blocks can thus be bounded by
$ {\ell \choose \frac{3\ell}{4}}\cdot 2^{- (C_1\log_d n)} \leq 2^\ell\cdot 2^{- (C_1\log_d n)}\leq 2^{- (C_1\log_d n)/2} = \exp(- \Omega(\log_d n)).$
\end{proof}

By a union bound, we get that there exists a suitable constant $C'_1$ such that $\Pr[\E_1\wedge \E_2] \ge (1-\frac{1}{2}\exp(-C'_1 \log_d n))$.
Observe that, when $\E_2$ occurs, then there are at least $\ell/4$ blocks which are visited by both $q_1$ and $q_2$.
If $\E_1$ occurs as well, then for each such block, $\sigma_X$ pays a cost of least $t$ since it visits a vertex in $X_1$ followed by a vertex in $X_2$, or vice-versa, and these vertices are at least $t$ apart.
So if both $\E_1$ and $\E_2$ occur, the cost of $\sigma_X$ is at least $t\ell/4 = \Omega(\log^2 n)$, since $d$ is a constant. Using the fact that $\opt_{TSP}(X) = O(\log n)$, we get that,
$\Pr_{X\leftarrow \D'}[c(\sigma_X) = o(\log n)\opt_{TSP}(X)] \le \frac{1}{2}\exp(-C'_1 \log_d n)$.
We choose the constant $C_0 := C'_1/\log d$ and set $\eps_0 := 2C'_1$; observe that we have $\eps_0|X| \le \frac{\eps_0\log n}{2\log d}= C_0\log n$. This ends the description of $\D'$  for which \eqref{eq:tsp} holds.
\end{proofof}

\vspace{-3mm}

\section{Strong Universal Lower Bounds imply Privacy Lower Bounds}
\label{sec:univpriv}
\vspace{-2mm}
Suppose $\Pi$ is a minimization problem whose instances
are indexed as tuples $(I,X)$. The first component $I$ represents the part of the input
that is accessible to the algorithm (and is public); for instance, in the Steiner tree
and the TSP example, this is the metric space $(V,c)$ along with the identity of the root.
The second component $X$ is the part of the input which is either unknown beforehand,
or corresponds to the private input. We assume that $X$ is a subset of some finite universe
$U = U(I)$.
In the Steiner tree and TSP example, $X$ is the set of terminals which
is a subset of all the vertices. An instance $(I,X)$ has a set of feasible solutions
$\script{S}(I,X)$, or simply $\script{S}(X)$ when $I$ is clear from context, and
let $\script{S} :=  \bigcup_{X\subseteq U}\script{S}(X)$.
In the case of Steiner trees, $\script{S}(X)$ is the collection of rooted
trees containing $X$; in the case of TSP it is the set of tours spanning $X\cup r$.
Every solution $S\in \script{S}$ has an
associated cost $c(S)$, and $\opt(X)$ denotes the solution of minimum
cost in $\script{S}(X)$.

We assume that the solutions to instances of $\Pi$ have the following {\em projection} property. Given any solution $S\in \script{S}(X)$ and any $X'\subseteq X$,
$S$ induces a unique solution in $\script{S}(X')$, denoted by $\pi_{X'}(S)$. For instance,
in case of the Steiner tree problem, a rooted tree spanning vertices of $X$ maps to the unique
minimal rooted tree spanning $X'$. Similarly, in the TSP, an ordering of vertices in $X$
maps to the induced ordering of $X'$. In this framework, we now define approximate
universal and differentially private algorithms.
%

An {\em $\alpha$-approximate universal algorithm} for $\Pi$ takes input $I$ and returns a distribution $\D$ over solutions
in $\script{S}(U)$
with the property that for any $X\subseteq U$, ${\bf E}_{S\leftarrow \D}[c(\pi_{X}(S))] \le \alpha\cdot \opt(I,X)$. An {\em $\alpha$-approximate differentially private algorithm} with {\em privacy parameter $\eps$} for $\Pi$
takes as input $(I,X)$ and returns a distribution $\D_X$ over solutions in
$\bigcup_{Y\supseteq X} \script{S}(Y)$ that satisfies the following two
properties. First, for all $(I,X)$, ${\bf E}_{S\leftarrow \D_X}[c(\pi_X(S))]\le \alpha\cdot \opt(I,X)$.
Second, for any set of solutions $\script{F}$ and for any pair of sets $X$ and $X'$ with symmetric
difference exactly $1$, we have
$$\exp(-\eps) \cdot  \Pr_{S\leftarrow \D_{X'}}[S\in \script{F}] \le \Pr_{S\leftarrow \D_{X}}[S\in \script{F}] \le \exp(\eps)\cdot \Pr_{S\leftarrow \D_{X'}}[S\in \script{F}]$$
\noindent
It is easy to see that any $\alpha$-approximate universal algorithm is also an $\alpha$-approximate differentially private algorithm
with privacy parameter $\eps = 0$; the distribution $\D_X := \D$ for every $X$ suffices.
We now show a converse relation: lower bounds for universal algorithms with a certain additional property imply lower bounds for private algorithms as well. We make this precise.

Fix $\rho:[n]\to [0,1]$ to be a non-increasing function.
We say that an {\em $(\alpha,\rho)$ lower bound} holds for universal algorithms if there
exists $I$ with the following property. Given any distribution $\script{D}$ on $\script{S}(U)$,
there exists a subset $X\subseteq U$ such that
\begin{equation}\label{eq:rhomu}
\Pr_{S\leftarrow \script{D}} [c(\pi_X(S)) \le \alpha \cdot \opt(I,X)] ~ \le ~ \rho(|X|)
\end{equation}
We say that the set $X$ achieves the $(\alpha,\rho)$ lower bound.
It is not hard to see that when $\rho$ is a constant function bounded away from $1$, an
$(\alpha,\rho)$ lower bound
is equivalent to an $\Omega(\alpha)$ lower bound on universal algorithms.

\begin{theorem}\label{thm:univpriv}
Suppose there exists a $(\alpha,\rho)$ lower bound for universal algorithms for a problem $\Pi$. Then any $\eps$-private
algorithm for $\Pi$ with $\eps \le \eps_0 := \inf_X \frac{1}{|X|}\ln\left(\frac{1}{2\rho(|X|)}\right)$ has an approximation factor of $\Omega(\alpha)$.
\end{theorem}
\begin{proof}
Let $I$ be an instance that induces the $(\alpha,\rho)$ lower bound.
Consider the output of a differentially private algorithm $\script{A}$ with privacy parameter $\eps< \eps_0$, on the input pair $(I,\emptyset)$.
Let $\D$ be the distribution on the solution set $\script{S}$. We first claim that
all $S$ in the support of $\D$ lie in $\script{S}(U)$. Suppose not and suppose there is a solution
$S\in \script{S}(Z)\setminus \script{S}(U)$, for some $Z\subset U$, which is returned with non-zero probability.
By the definition of differential privacy, this solution must be returned with non-zero probability when $\script{A}$
is run with $(I,U)$, contradicting feasibility since $S\notin\script{S}(U)$.

Thus, $\D$ can be treated as a universal solution for $\Pi$. Let $X$ be the set which achieves the $(\alpha,\rho)$
lower bound for $\D$, and let $\script{F} := \{S\in \script{S}(X): c(S) \le \alpha\cdot \opt(I,X)\}$.
By the definition of the lower bound,
we know that $\Pr_{S\leftarrow \script{D}} [S\in \script{F}] ~ \le ~ \rho(|X|)$.
Let $\D'$ be the output of the algorithm $\script{A}$ when the input is $(I,X)$. By definition of differential privacy,
$\Pr_{S\leftarrow \script{D'}} [S\in \script{F}] ~ \le ~ \exp(\eps\cdot|X|)\cdot \rho(|X|) \le 1/2$, from the choice
of $\eps$. This shows a lower bound on the approximation factor of any differential private algorithm
for $\Pi$ with parameter $\eps < \eps_0$.
\end{proof}

\bibliographystyle{plain}
\bibliography{huge}
\appendix
\section{Upper Bounds on Universal Algorithms}\label{sec:prelims}
\label{sec:upper_bounds}
\vspace{-2mm}

In Section \ref{sec:prevwork}, we mention that there exist $O(\log n)$-approximate algorithms for the
universal Steiner tree problem and the universal TSP. The Steiner tree result
follows from the results of probabilistic embedding general metrics into tree metrics; this was remarked by Jia et al.~\cite{JLNRS05}
and Gupta et al.~\cite{GLMRT10}. The TSP result follows from the observation that any $\alpha$-approximate
Steiner tree algorithm implies an $2\alpha$-approximate universal TSP algorithm; this follows from a standard argument of
obtaining a tour of from a tree of at most double the cost by performing a depth first traversal.
This was noted by Schalekamp and Shmoys \cite{SS08}.
We remark here that  this reduction {\em does not} hold when the Steiner tree algorithm is allowed to return a collection of paths; in particular, our lower bound for universal TSP (Theorem \ref{thm:univtsp}) {\em does not imply} the lower bound for universal Steiner tree algorithms  which return path collections (Theorem \ref{thm:univst-paths}). For completeness, we give short proofs of the above
two observations.

\indent
Given a metric space $(V,c)$ and any spanning tree $T$ of $V$,
let $c_T(u,v)$, for any two vertices $u,v$, be
the cost of all the edges in the unique path connecting $u$ and $v$ in $T$.
Given a distribution $\script{D}$ on spanning trees,  define the
stretch of a pair $(u,v)$ to be $\frac{{\bf E}_{T\leftarrow \script{D}}[c_T(u,v)]}{c(u,v)}$.
The stretch of $\script{D}$ is $\max_{(u,v)\in V\times V} \mbox{stretch}(u,v)$.
The following connects the stretch and the performance of this algorithm.

\begin{theorem}
Suppose there exists a distribution $\script{D}$ on spanning trees that has stretch
at most $\alpha$. Then the distribution gives an $\alpha$-approximation for the
universal steiner tree problem.
\end{theorem}
\vspace{-2mm}
\begin{proof}
Fix any set of terminals $X$. Let $T^*$ be the tree which attains value $\opt_{ST}(X)$.
Let the support of $\D$ be $(T_1,\ldots,T_\ell)$ with $\pi_i$ being the probability
of $T_i$. For every edge $(u,v)\in T^*$, let $p_i(u,v)$ be the unique $u,v$ path in $T_i$.
Note that $\bigcup_{(u,v)\in T^*} p_i(u,v)$ is a sub-tree of $T_i$ which connects $X$, and thus,
$c(T_i[X]) \le c\left(\bigcup_{(u,v)\in T^*} p_i(u,v)\right) \le \sum_{(u,v)\in T^*} c_{T_i}(u,v)$.
Thus, the expected cost of the universal Steiner tree algorithm is
$$\sum_{i=1}^\ell \pi_ic(T_i[X]) \le \sum_{i=1}^\ell \pi_i \ \sum_{(u,v)\in T^*} c_{T_i}(u,v) = \sum_{(u,v)\in T^*}\sum_{i=1}^\ell \pi_ic_{T_i}(u,v)
=
\sum_{(u,v)\in T^*} {\bf E}_{T\leftarrow \script{D}} [c_T(u,v)]
\le \alpha \cdot \opt_{ST}(X)$$
\hfill
\end{proof}

\noindent
It is known by the results of Fakcharoenphol et al. \cite{FRT03} that for any $n$ vertex metric $(V,c)$
one can find a a distribution $\D$ with stretch $O(\log n)$.\footnote{Strictly speaking, the algorithms of \cite{FRT03} do not
return a distribution on spanning trees, but rather a distribution on what are known as hierarchically well-separated trees. However, it is known
that with another constant factor loss, one can obtain an embedding onto spanning trees of $V$ as well. See Section 5 of the
paper \cite{KRS01}, for instance.}
This gives us the following corollary.
\begin{corollary}\label{cor:univst}
There is an $O(\log n)$-approximate universal Steiner tree algorithm.
\end{corollary}
\begin{theorem}
An $\alpha$-approximate universal Steiner tree algorithm implies a $2\alpha$-approximate universal TSP algorithm.
\end{theorem}
\begin{proof}
Suppose the $\alpha$-approximate universal Steiner tree algorithm returns a distribution $\script{D}$ on spanning trees.
For each tree $T$ in the support of $\D$, consider the ordering $\sigma$ of the vertices obtained
by performing a depth-first traversal of the tree. This induces a distribution on orderings, and thus a
universal TSP algorithm. We claim this is $2\alpha$-approximate.
Fix any subset $X\subseteq V$ and let $T[X]$ be the unique minimal tree of $T$ which spans
$X\cup r$. Let $\sigma'$ be the ordering of the vertices in $T[X]$ obtained on performing a depth-first
traversal of $T[X]$.
\begin{claim}
The order in which $\sigma'$ visits vertices of $X$ is the same order in which $\sigma$ visits them.
\end{claim}
\begin{proof}
$T[X]$ is obtained from $T$ by deleting a collection of sub-trees from $T$. Note that all the vertices
of any sub-tree appear contiguously
in any depth-first traversal order - this is because once the depth first
traversal visits a vertex $v$, it traverses all vertices in the sub-tree of $v$ before moving on to any
other vertex not in the sub-tree of $v$. Therefore, deleting a sub-tree of $T$ and performing
a depth first traversal only removes a contiguous piece
in the ordering $\sigma$. The ordering of the remaining
vertices is left unchanged.
\end{proof}

To complete the proof, we use the fact that if $\sigma$ is the depth first traversal order of any tree $T$,
then $c(\sigma_{V(T)}) \le 2c(E(T))$ - this is because any edge of $T$ is traversed at most twice once in the forward
direction and one reverse. Thus, $c(\sigma(X)) = c(\sigma'(X)) \le 2c(T[X]) \le 2\alpha\cdot \opt_{ST}(X) \le 2\alpha\cdot \opt_{TSP}(X)$,
where the last inequality uses that the tour of $X$ contains a Steiner tree of $X$. \end{proof}
\begin{corollary}
There exists an $O(\log n)$ approximation for the universal TSP.
\end{corollary}

\end{document}